\documentclass[aps,showpacs,twocolumn]{revtex4}
\usepackage{epsfig}

\begin{document}

\title{Possible $H$-like dibaryon states with heavy
quarks\footnote{Corresponding author: jlping@njnu.edu.cn (J.L. Ping)}}

\author{Hongxia Huang$^{a}$, Jialun Ping$^a$, and Fan Wang$^b$}

\affiliation{$^a$Department of Physics, Nanjing Normal University,
Nanjing 210097, P.R. China}

\affiliation{$^b$Department of Physics, Nanjing University,
Nanjing 210093, P.R. China}

\begin{abstract}
Possible $H$-like dibaryon states $\Lambda_{c}\Lambda_{c}$ and
$\Lambda_{b}\Lambda_{b}$ are investigated within the framework of
quark delocalization color screening model. The results
show that the interaction between two $\Lambda_{c}$'s is repulsive,
so it cannot be bound state by itself. However, the strong attraction in
$\Sigma_{c}\Sigma_{c}$ and $\Sigma^{*}_{c}\Sigma^{*}_{c}$ channels
and the strong channel coupling, due to the central interaction of one-gluon-exchange
and one-pion-exchange, among $\Lambda_{c}\Lambda_{c}$, $\Sigma_{c}\Sigma_{c}$
and $\Sigma^{*}_{c}\Sigma^{*}_{c}$ push the energy of system below the threshold
of $\Lambda_{c}\Lambda_{c}$ by $22$ MeV. The
corresponding system $\Lambda_{b}\Lambda_{b}$ has the similar
properties as that of $\Lambda_{c}\Lambda_{c}$ system, and a bound
state is also possible in $\Lambda_{b}\Lambda_{b}$ system.
\end{abstract}


\maketitle

\setcounter{totalnumber}{5}

\section{\label{sec:introduction}Introduction}

The $H$ dibaryon, a six quark ($uuddss$) state corresponding
asymptotically to a bound $\Lambda\Lambda$ system, was first
proposed by Jaffe in 1977~\cite{Jaffe}. This hypothesis initiated
a worldwide activity of theoretical studies and experimental
searches for dibaryon states~\cite{Jaffe2}. In 1987, M. Oka $et~
al.$ claimed that a sharp resonance appears in $^{1}S_{0}$
$\Lambda\Lambda$ scattering at $E_{c.m.}=26.3$ MeV, which might
correspond to the $H$ dibaryon state~\cite{Oka}. Moreover, M. Oka
also proposed several $J^{P}=2^{+}$ dibaryons in the quark cluster
model without meson exchange~\cite{Oka2}. Despite numerous claims,
no dibaryon candidate has been confirmed experimentally so far.
Recntly, the interest in the $H$ dibaryon have been revived by
lattice QCD calculations of
different collaborations, NPLQCD~\cite{NPL} and HALQCD~\cite{HAL}.
These two groups reported that the $H$ particle is indeed a bound
state at pion masses larger than the physical ones. Then, Carames
and Valcarce examined the $H$ dibaryon within a chiral constituent
quark model and obtained a bound $H$ dibaryon with binding energy
$B_{H}=7$ MeV~\cite{Carames}.

Understanding the hadron-hadron interactions and searching exotic quark
states are important topics in temporary hadron physics.
Recently observed many near-threshold charmonium-like states, such
as $X(3872)$, $Y(3940)$, and $Z^{+}(4430)$, triggered lots of
studies on the molecule-like bound states containing heavy quark
hadrons. Such a study may help us to understand further the
hadron-hadron interactions. In the heavy quark sector, the large
masses of the heavy baryons reduce the kinetic of the system,
which makes it easier to form bound states. One may wonder whether
a $H$-like dibaryon state $\Lambda_{c}\Lambda_{c}$ exist or not.

In particular, the deuteron is a loosely bound state of a proton
and a neutron, which may be regarded as a hadronic molecular
state. The possibility of existing deuteron-like states, such as
$N\Sigma_{c}$, $N\Xi^{'}_{c}$, $N\Xi_{cc}$, $\Xi\Xi_{cc}$ and so
on, were investigated by several realistic phenomenological
nucleon-nucleon interaction models~\cite{Fromel,Julia}. The
$N\Lambda_{c}$ system and relevant coupled channel effects were
both studied on hadron level~\cite{Liu} and on quark
level~\cite{Huang}. However, some different results were obtained
by these two methods. On hadron level~\cite{Liu}, it is found that
molecular bound states of $N\Lambda_{c}$ are plausible in both the
one-pion-exchange potential model and the one-boson-exchange
potential model. On quark level~\cite{Huang}, our group found the
attraction between $N$ and $\Lambda_{c}$ is not strong enough to
form any $N\Lambda_{c}$ bound state within our quark
delocalization color screening model (QDCSM). Whereas the
attraction between $N$ and $\Sigma_{c}$ is strong enough to form a
bound state $N\Sigma_{c}(^{3}S_{1})$, it becomes a resonance state
by coupling to the open $N\Lambda_{c}$ $D-$wave channels. We also
explored the corresponding states $N\Lambda_{b}$, $N\Sigma_{b}$
and the similar properties as that of states $N\Lambda_{c}$,
$N\Sigma_{c}$ were obtained. Recently, the possible
$\Lambda_{c}\Lambda_{c}$ molecular state was studied in the
one-boson-exchange potential model~\cite{Lee} and in the
one-pion-exchange potential model~\cite{Meguro} on hadron level.
Different results were obtained by these two models. The
$\Lambda_{c}\Lambda_{c}$ does not exist in the former model,
whereas the molecular bound state of $\Lambda_{c}\Lambda_{c}$ is
possible in the later model. So the quark level study of the
$\Lambda_{c}\Lambda_{c}$ system is interesting and necessary.

The quark delocalization color screening model (QDCSM) was
developed in the 1990s with the aim of explaining the similarities
between nuclear and molecular forces~\cite{QDCSM0}. The model
gives a good description of $NN$ and $YN$ interactions and the
properties of deuteron~\cite{QDCSM1}. It is also employed to
calculate the baryon-baryon scattering phase shifts in the
framework of the resonating group method (RGM), and the dibaryon
candidates are also studied with this model~\cite{QDCSM2,QDCSM3}.
Recent study also show that the intermediate-range attraction
mechanism in the QDCSM, quark delocalization and color screening,
is an alternative mechanism for the $\sigma$-meson exchange in the
most common quark model, the chiral quark
model~\cite{QDCSM2,QDCSM3}. In the frame of QDCSM, the $H$
dibaryon were also obtained~\cite{QDCSM4}. Therefore, it is very
interesting to investigate whether a $H$-like dibaryon state
$\Lambda_{c}\Lambda_{c}$ exist or not in QDCSM.

In present work, QDCSM is employed to study the properties of
$\Lambda_{c}\Lambda_{c}$ systems. the channel-coupling effect
of $\Sigma_{c}\Sigma_{c}$, $\Sigma_{c}\Sigma^{*}_{c}$,
$\Sigma^{*}_{c}\Sigma^{*}_{c}$ and $N\Xi_{cc}$ are included.
Our purpose is to understand the interaction properties of the
$\Lambda_{c}\Lambda_{c}$ system and to see whether an $H$-like
dibaryon state $\Lambda_{c}\Lambda_{c}$ exist or not. Extension
of the study to the bottom case is also interesting and is performed
here. The structure of this paper is as follows.
After the introduction, we present a brief introduction of the
quark models used in section II. Section III devotes to the
numerical results and discussions. The summary is shown in the
last section.

\section{The quark delocalization color screening
model (QDCSM)}

The detail of QDCSM used in the present work can be found  in the
references~\cite{QDCSM0,QDCSM1,QDCSM2,QDCSM3}. Here, we
just present the salient features of the model. The model
Hamiltonian is:
\begin{widetext}
\begin{eqnarray}
H &=& \sum_{i=1}^6 \left(m_i+\frac{p_i^2}{2m_i}\right) -T_c
+\sum_{i<j} \left[ V^{G}(r_{ij})+V^{\chi}(r_{ij})+V^{C}(r_{ij})
\right],
 \nonumber \\
V^{G}(r_{ij})&=& \frac{1}{4}\alpha_{s_{ij}} {\mathbf \lambda}_i
\cdot {\mathbf \lambda}_j
\left[\frac{1}{r_{ij}}-\frac{\pi}{2}\left(\frac{1}{m_{i}^{2}}+\frac{1}{m_{j}^{2}}+\frac{4{\mathbf
\sigma}_i\cdot {\mathbf\sigma}_j}{3m_{i}m_{j}}
 \right)
\delta(r_{ij})-\frac{3}{4m_{i}m_{j}r^3_{ij}}S_{ij}\right],
\nonumber \\
V^{\chi}(r_{ij})&=& \frac{1}{3}\alpha_{ch}
\frac{\Lambda^2}{\Lambda^2-m_{\chi}^2}m_\chi \left\{ \left[
Y(m_\chi r_{ij})- \frac{\Lambda^3}{m_{\chi}^3}Y(\Lambda r_{ij})
\right]
{\mathbf \sigma}_i \cdot{\mathbf \sigma}_j \right.\nonumber \\
&& \left. +\left[ H(m_\chi r_{ij})-\frac{\Lambda^3}{m_\chi^3}
H(\Lambda r_{ij})\right] S_{ij} \right\} {\mathbf F}_i \cdot
{\mathbf F}_j, ~~~\chi=\pi,K,\eta \\
V^{C}(r_{ij})&=& -a_c {\mathbf \lambda}_i \cdot {\mathbf
\lambda}_j [f(r_{ij})+V_0], \nonumber
\\
 f(r_{ij}) & = &  \left\{ \begin{array}{ll}
 r_{ij}^2 &
 \qquad \mbox{if }i,j\mbox{ occur in the same baryon orbit} \\
  \frac{1 - e^{-\mu_{ij} r_{ij}^2} }{\mu_{ij}} & \qquad
 \mbox{if }i,j\mbox{ occur in different baryon orbits} \\
 \end{array} \right.
\nonumber \\
S_{ij} & = &  \frac{{\mathbf (\sigma}_i \cdot {\mathbf r}_{ij})
({\mathbf \sigma}_j \cdot {\mathbf
r}_{ij})}{r_{ij}^2}-\frac{1}{3}~{\mathbf \sigma}_i \cdot {\mathbf
\sigma}_j. \nonumber
\end{eqnarray}
\end{widetext}
Where $S_{ij}$ is quark tensor operator, $Y(x)$, $H(x)$ and $G(x)$
are standard Yukawa functions~\cite{Valcarce}, $T_c$ is the
kinetic energy of the center of mass, $\alpha_{ch} $ is the chiral
coupling constant, determined as usual from the $\pi$-nucleon
coupling constant. All other symbols have their usual meanings.
Here, a phenomenological color screening confinement potential is
used, and $\mu_{ij}$ is the color screening parameter. For the
light-flavor quark system, it is determined by fitting the
deuteron properties, $NN$ scattering phase shifts, $N\Lambda$ and
$N\Sigma$ scattering phase shifts, respectively, with
$\mu_{uu}=1.2$, $\mu_{us}=0.3$, $\mu_{ss}=0.08$, satisfying the
relation $\mu^{2}_{us}=\mu_{uu}*\mu_{ss}$~\cite{QDCSM3}. When
extending to the heavy quark case, there is no experimental data
available, so we take it as a common parameter. In the present
work, we take $\mu_{cc}=0.001$ and $\mu_{uc}=0.0346$, also satisfy
the relation $\mu^{2}_{uc}=\mu_{uu}*\mu_{cc}$. All the other
parameters are taken from~\cite{Huang}.

The quark delocalization in QDCSM is realized by specifying the
single particle orbital wave function of QDCSM as a linear
combination of left and right Gaussians, the single particle
orbital wave functions used in the ordinary quark cluster model,
\begin{eqnarray}
\psi_{\alpha}(\mathbf{s}_i ,\epsilon) & = & \left(
\phi_{\alpha}(\mathbf{s}_i)
+ \epsilon \phi_{\alpha}(-\mathbf{s}_i)\right) /N(\epsilon), \nonumber \\
\psi_{\beta}(-\mathbf{s}_i ,\epsilon) & = &
\left(\phi_{\beta}(-\mathbf{s}_i)
+ \epsilon \phi_{\beta}(\mathbf{s}_i)\right) /N(\epsilon), \nonumber \\
N(\epsilon) & = & \sqrt{1+\epsilon^2+2\epsilon e^{-s_i^2/4b^2}}. \label{1q} \\
\phi_{\alpha}(\mathbf{s}_i) & = & \left( \frac{1}{\pi b^2}
\right)^{3/4}
   e^{-\frac{1}{2b^2} (\mathbf{r}_{\alpha} - \mathbf{s}_i/2)^2} \nonumber \\
\phi_{\beta}(-\mathbf{s}_i) & = & \left( \frac{1}{\pi b^2}
\right)^{3/4}
   e^{-\frac{1}{2b^2} (\mathbf{r}_{\beta} + \mathbf{s}_i/2)^2}. \nonumber
\end{eqnarray}
Here $\mathbf{s}_i$, $i=1,2,...,n$ are the generating coordinates,
which are introduced to expand the relative motion
wavefunction~\cite{QDCSM1}. The mixing parameter
$\epsilon(\mathbf{s}_i)$ is not an adjusted one but determined
variationally by the dynamics of the multi-quark system itself.
This assumption allows the multi-quark system to choose its
favorable configuration in the interacting process. It has been
used to explain the cross-over transition between hadron phase and
quark-gluon plasma phase~\cite{Xu}.

\section{The results and discussions}

Here, we perform a dynamical investigation of the
$\Lambda_{c}\Lambda_{c}$ system with $IJ^{P}=00^{+}$ in the QDCSM.
The channel coupling effects are also considered. The labels of
all coupled channels are listed in Table \ref{channels}.

\begin{widetext}
\begin{center}
\begin{table}[h]
\caption{The $\Lambda_{c}\Lambda_{c}$ and $\Lambda_{b}\Lambda_{b}$
states and the channels coupled to them.}
\begin{tabular}{lcccccccc}
\hline \hline
 Channels & 1 & ~~~2 & ~~~3 & ~~~4 & ~~~5 & ~~~6 & ~~~7  \\ \hline
 $J^{P}=0^{+}$~~~ & $\Sigma_{c}\Sigma_{c}(^{1}S_{0})$ & ~~~$N\Xi_{cc}(^{1}S_{0})$
                  & ~~~$\Lambda_{c}\Lambda_{c}(^{1}S_{0})$ & ~~~$\Sigma^{*}_{c}\Sigma^{*}_{c}(^{1}S_{0})$
                  & ~~~$N\Xi^{*}_{cc}(^{5}D_{0})$  & ~~~$\Sigma_{c}\Sigma^{*}_{c}(^{5}D_{0})$
                  & ~~~$\Sigma^{*}_{c}\Sigma^{*}_{c}(^{5}D_{0})$  \\  \hline
 $J^{P}=0^{+}$~~~ & $\Sigma_{b}\Sigma_{b}(^{1}S_{0})$ & ~~~$N\Xi_{bb}(^{1}S_{0})$
                  & ~~~$\Lambda_{b}\Lambda_{b}(^{1}S_{0})$ & ~~~$\Sigma^{*}_{b}\Sigma^{*}_{b}(^{1}S_{0})$
                  & ~~~$N\Xi^{*}_{bb}(^{5}D_{0})$  & ~~~$\Sigma_{b}\Sigma^{*}_{b}(^{5}D_{0})$
                  & ~~~$\Sigma^{*}_{b}\Sigma^{*}_{b}(^{5}D_{0})$  \\  \hline
  \hline
\end{tabular}
\label{channels}
\end{table}
\end{center}
\end{widetext}

\begin{figure*}
\epsfxsize=5.5in \epsfbox{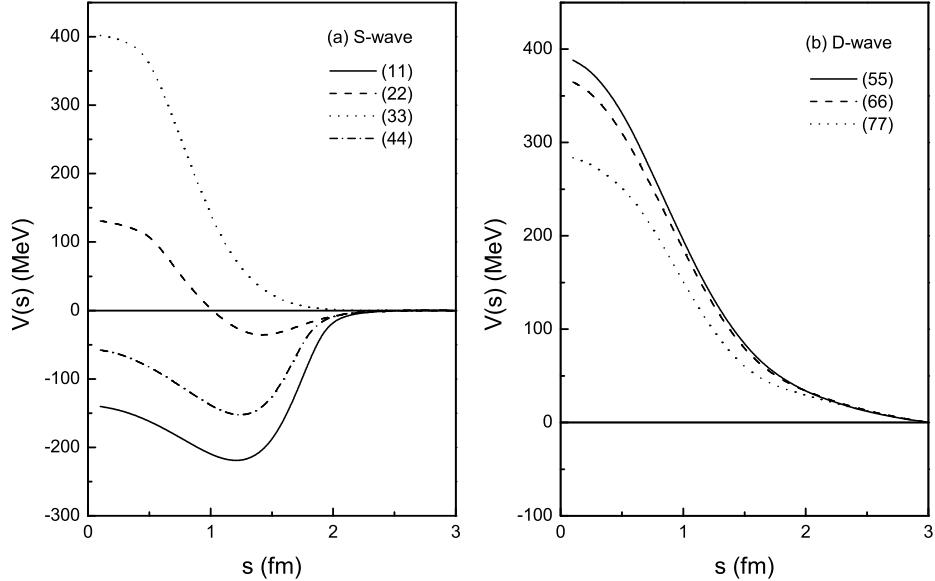} 
\vspace{-0.4in}

\caption{The potentials of different channels for the
$J^{P}=0^{+}$ case of the $\Lambda_{c}\Lambda_{c}$ system.}
\end{figure*}

Because an attractive potential is necessary for forming bound
state or resonance, we first calculate the effective potentials of
all the channels listed in Table \ref{channels}. The effective
potential between two colorless clusters is defined as,
$V(s)=E(s)-E(\infty)$, where $E(s)$ is the diagonal matrix element
of the Hamiltonian of the system in the generating coordinate. The
effective potentials of the $S$-wave and $D$-wave channels are
shown in Fig. 1(a) and (b) respectively. From Fig. 1(a), we can
see that the potentials are attractive for the $^{1}S_{0}$
channels $\Sigma_{c}\Sigma_{c}$, $N\Xi_{cc}$ and
$\Sigma^{*}_{c}\Sigma^{*}_{c}$. While for the channel
$\Lambda_{c}\Lambda_{c}$, the potential is repulsive and so no
bound state can be formed in this single channel. However, the
attractions of $\Sigma_{c}\Sigma_{c}$ channel and
$\Sigma^{*}_{c}\Sigma^{*}_{c}$ channel are very large, the channel coupling effects
of $\Sigma_{c}\Sigma_{c}$ and $\Sigma^{*}_{c}\Sigma^{*}_{c}$ to
$\Lambda_{c}\Lambda_{c}$ will push the energy of $\Lambda_{c}\Lambda_{c}$
downward, it is possible to from a bound state. For the
$^{5}D_{0}$ channels shown in Fig. 1(b), the potentials are all
repulsive.

In order to see whether or not there is any bound state, a dynamic
calculation is needed. Here the RGM equation is employed. Expanding
the relative motion wavefunction
between two clusters in the RGM equation by gaussians, the
integro-differential equation of RGM can be reduced to algebraic
equation, the generalized eigen-equation. The energy of the system
can be obtained by solving the eigen-equation. In the calculation,
the baryon-baryon separation ($|\mathbf{s}_n|$) is taken to be less
than 6 fm (to keep the matrix dimension manageably small).

The single channel calculation shows that the energy of
$\Lambda_{c}\Lambda_{c}$ is above its threshold, the sum of masses
of two $\Lambda_{c}$'s. It is reasonable, because the interaction
between the two $\Lambda_{c}$'s is repulsive as mentioned above.
For $N\Xi_{cc}$ channel, the attraction is too weak to tie the
two particles together, so it is also unbound. At the same time, due to
the stronger attraction, the energies of $\Sigma_{c}\Sigma_{c}$
and $\Sigma^{*}_{c}\Sigma^{*}_{c}$ are below their corresponding
thresholds. The binding energy of $\Sigma_{c}\Sigma_{c}$ and
$\Sigma^{*}_{c}\Sigma^{*}_{c}$ states are listed in Table
\ref{bound_c}, in which '$ub$' means unbound. For the $^{5}D_{0}$
channels, they are all unbound since the potentials are all
repulsive, so we leave them out from Table \ref{bound_c}. We also
do a channel-coupling calculation and a bound state, which energy is below
the threshold of $\Lambda_{c}\Lambda_{c}$, is obtained.
The binding energy is also shown in Table \ref{bound_c} under the head
'$c.c.$'. There are several features which are discussed below.

\begin{table}[ht]
\caption{The binding energy of every $^{1}S_{0}$ channels of
$\Lambda_{c}\Lambda_{c}$ system and with channel coupling
($c.c.$).}
\begin{tabular}{lccccccc}
\hline \hline
 Channels & ~~$\Sigma_{c}\Sigma_{c}$~~  & ~~$N\Xi_{cc}$~~ & ~~$\Lambda_{c}\Lambda_{c}$~~ & ~~$\Sigma^{*}_{c}\Sigma^{*}_{c}$~~ & ~~$c.c.$~~\\
\hline
 B.E.(MeV) & $-157$ & $ub$  & $ub$ & $-91$ & $-22$    \\
\hline \hline
\end{tabular}
\label{bound_c}
\end{table}

\begin{figure*}
\epsfxsize=5.5in \epsfbox{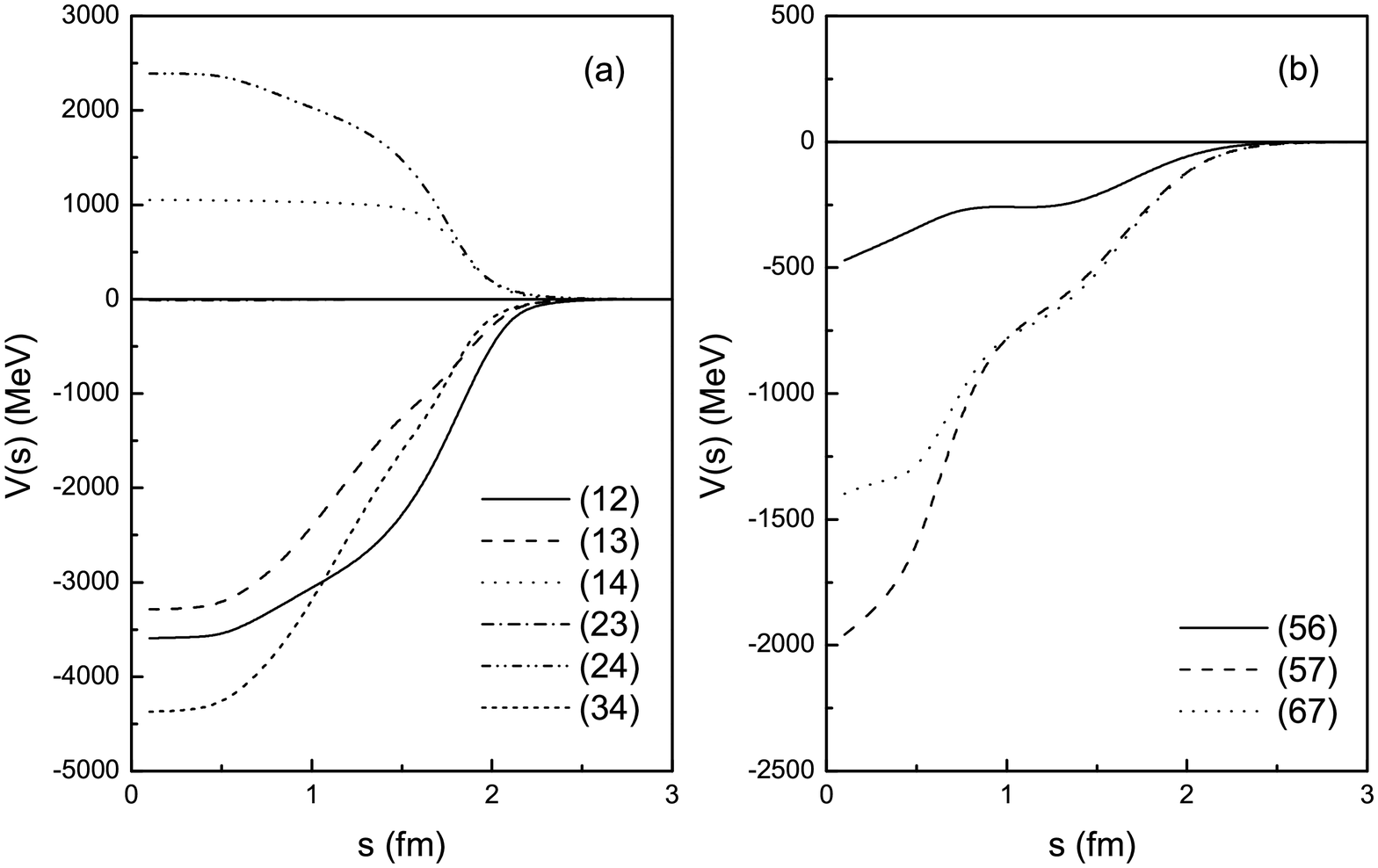} 
\vspace{-0.4in}

\caption{The transition potentials of (a): $S$-wave channels and
(b): $D$-wave channels for the $J^{P}=0^{+}$ case of the
$\Lambda_{c}\Lambda_{c}$ system.}
\end{figure*}

\begin{figure*}
\epsfxsize=5.5in \epsfbox{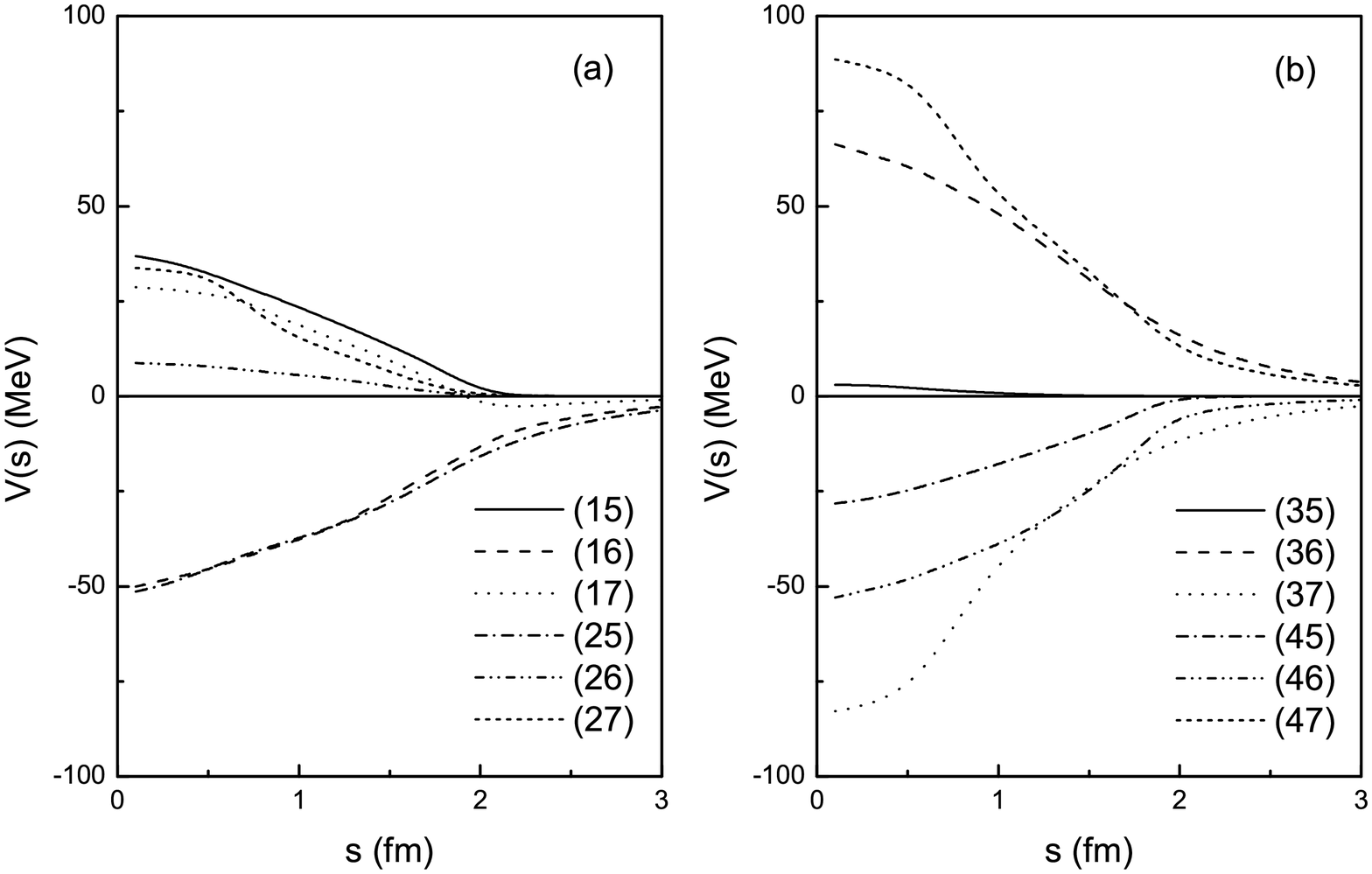} 
\vspace{-0.4in}

\caption{The transition potentials of $S-D$ wave channels for the
$J^{P}=0^{+}$ case of the $\Lambda_{c}\Lambda_{c}$ system.}
\end{figure*}

First, the individual $S$-wave $\Lambda_{c}\Lambda_{c}$ channel is
unbound in our quark level calculation, which is consistent
with the conclusion on the hadron level~\cite{Lee,Meguro}. For
other individual channels, there are some different results. In
Ref.~\cite{Meguro}, the calculation shows all the individual
channels are unbound and in Ref.~\cite{Lee}, $\Sigma_{c}\Sigma_{c}$ is
 bound. While in our quark level calculation, the
individual $\Sigma_{c}\Sigma_{c}$ and
$\Sigma^{*}_{c}\Sigma^{*}_{c}$ are deeply bound.

Secondly, by taking into account the channel-coupling effect, a
bound state is obtained for the $\Lambda_{c}\Lambda_{c}$ system,
which is also consistent with the conclusion on the hadron
level~\cite{Meguro}. However, the channel-coupling effect is
different between our quark level calculation and their hadron
level calculation. In Ref.~\cite{Meguro}, the coupling of
$\Lambda_{c}\Lambda_{c}$ to the $D$-wave channels $\Sigma_{c}\Sigma^{*}_{c}$
and $\Sigma^{*}_{c}\Sigma^{*}_{c}$ are crucial in binding two
$\Lambda_{c}$'s. This indicates the importance of the tensor force.
This conclusion is the same as their calculation of $N\Lambda_{c}$
system~\cite{Liu}. While in our quark level calculation, the
coupling between $\Lambda_{c}\Lambda_{c}$, $N\Xi_{cc}$,
$\Sigma_{c}\Sigma_{c}$ and $\Sigma^{*}_{c}\Sigma^{*}_{c}$ channels
is through the central force. The transition potentials of these
four channels are shown in Fig. 2(a). It is the strong coupling among
these channels that makes the $\Lambda_{c}\Lambda_{c}(^{1}S_{0})$ be
bound state. The transition potentials of three $D$-wave channels are
shown in Fig. 2(b).
To see the effects of tensor force, the transition potentials
for $S$ and $D$ wave channels are shown in Fig. 3(a) and (b).
From which one can see that the effects of tensor force are much small compared with
that of the central force. Thus the $S$ and $D$ wave channel-coupling effect is small
in out quark model calculation. This conclusion is
consistent with our calculation of $N\Lambda_{c}$
system~\cite{Huang}, in which the effect of the
$N\Sigma^{*}_{c}(^{5}D_{0})$ channel coupling to
$N\Lambda_{c}(^{1}S_{0})$ is very small.

Thirdly, the properties of the $\Lambda_{c}\Lambda_{c}$ system in
our quark model is similar to that of the $\Lambda\Lambda$ system.
Our group has calculated the $H$-dibaryon
before~\cite{QDCSM4}, in which the single channel $\Lambda\Lambda$
is unbound, when coupled to the channels $N\Xi$ and
$\Sigma\Sigma$, it becomes a bound state. Here, we extend our
model to study the heavy flavor dibaryons, we find it is possible
to form a bound state in the $\Lambda_{c}\Lambda_{c}$ system,
which it is a $H$-like dibaryon state.

In the previous discussion, the $\Lambda_{c}\Lambda_{c}$ system is
investigated and a $H$-like dibaryon state is found. Because of the
heavy flavor symmetry, we also extend the study to the bottom case
of $\Lambda_{b}\Lambda_{b}$ system. The numerical results for the
$N\Lambda_{b}$ system are listed in Figs. 4 and Table
\ref{bound_b}. The results are similar to the $\Lambda_{c}\Lambda_{c}$
system. From Table \ref{bound_b}, we also find there is also a $H$-like
dibaryon state in the $\Lambda_{b}\Lambda_{b}$ system in our quark model.

\begin{figure*}
\epsfxsize=5.5in \epsfbox{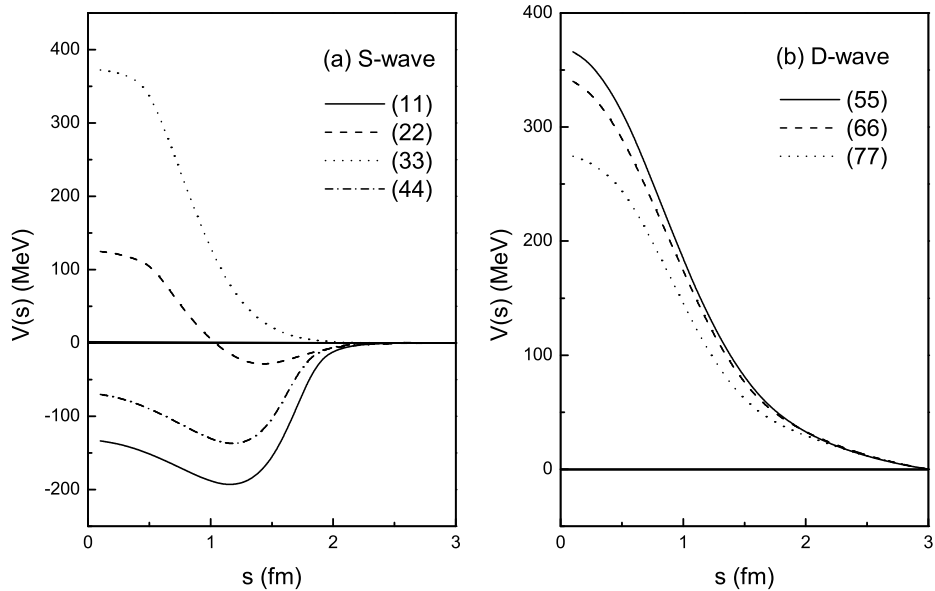} 
\vspace{-0.4in}

\caption{The potentials of different channels for the
$J^{P}=0^{+}$ case of the $\Lambda_{b}\Lambda_{b}$ system.}
\end{figure*}

\begin{table}[ht]
\caption{The binding energy of every $^{1}S_{0}$ channels of
$\Lambda_{b}\Lambda_{b}$ system and with channel coupling
($c.c.$).}
\begin{tabular}{lccccccc}
\hline \hline
 Channels & ~~$\Sigma_{b}\Sigma_{b}$~~  & ~~$N\Xi_{bb}$~~ & ~~$\Lambda_{b}\Lambda_{b}$~~ & ~~$\Sigma^{*}_{b}\Sigma^{*}_{b}$~~ & ~~$c.c.$~~\\
\hline
 B.E.(MeV) & $-162$ & $ub$  & $ub$ & $-79$ & $-19$    \\
\hline \hline
\end{tabular}
\label{bound_b}
\end{table}

\section{Summary}

In this work, we perform a dynamical calculation of the
$\Lambda_{c}\Lambda_{c}$ system with $IJ^{P}=00^{+}$ in the
framework of QDCSM. Our results show that the interaction between
two $\Lambda_{c}$'s is repulsive, so it cannot be a bound state by
itself. The attractions of $\Sigma_{c}\Sigma_{c}$ and
$\Sigma^{*}_{c}\Sigma^{*}_{c}$ channels are strong enough to bind
two $\Sigma_{c}$'s and two $\Sigma^*_{c}$'s together. It is
possible to form a $H$-like dibaryon state in the $\Lambda_{c}\Lambda_{c}$
system with the binding energy $22$ MeV in our quark model by
including the channel-coupling
effect.  This result is consistent with the result of the
calculation on the hadron level~\cite{Meguro}. However, the effect
of the channel coupling is different between these two approaches.
The role of the central force is much more important than the
tensor force in our quark level calculation, while in the
calculation on the hadron level~\cite{Meguro}, the tensor force is
shown to be important and the $D$-wave channels are crucial in
binding two $\Lambda_{c}$'s. Further investigation should be done
to understand the difference between the approaches on the hadron
level and the quark level. It will help us to understand the
quark-duality and exotic quark states.

Extension of the study to the bottom case has also been done. The
results of $\Lambda_{b}\Lambda_{b}$ system is similar to the
$\Lambda_{c}\Lambda_{c}$ system, and there exits a $H$-like
dibaryon state in the $\Lambda_{b}\Lambda_{b}$ system with a
binding energy of $19$ MeV in our quark model. On the experimental
side, finding the $H$-like dibaryon states $\Lambda_{c}\Lambda_{c}$
and $\Lambda_{b}\Lambda_{b}$ will be a challenging subject.

\end{document}